\documentclass[12pt,preprintnumbers]{revtex4}%
\usepackage[english]{babel}
\usepackage[T1]{fontenc}
\usepackage{bbm}
\usepackage{epsfig}
\usepackage{psfrag}
\usepackage{color}
\usepackage{amsmath}
\usepackage{amsfonts}
\usepackage{amssymb}
\usepackage{graphicx}%
\providecommand{\U}[1]{\protect\rule{.1in}{.1in}}
\newcommand{\be}{\begin{equation}}
\newcommand{\ee}{\end{equation}}
\newcommand{\bea}{\begin{eqnarray}}
\newcommand{\eea}{\end{eqnarray}}

\begin{document}
\title{Cosmological solutions with massive gravitons}
\author{Ali~H.~Chamseddine$^{1,2,3,4}$}
\author{Mikhail~S.~Volkov$^{2}$}
\affiliation{$^{1}$Physics Department, American University of Beirut, LEBANON}
\affiliation{$^{2}$Laboratoire de Math\'{e}matiques et Physique Th\'{e}orique CNRS-UMR
6083, Universit\'{e} de Tours, Parc de Grandmont, 37200 Tours, FRANCE}
\affiliation{$^{3}$LE STUDIUM, Loire Valley Institute for Advanced Studies, Tours and
Orleans, FRANCE}
\affiliation{$^{4}$I.H.E.S. F-91440 Bures-sur-Yvette, France}

\begin{abstract}
\vspace{1 cm} We present solutions describing spatially closed, open, or flat
cosmologies in the massive gravity theory within the recently proposed tetrad
formulation. We find that the effect of the graviton mass is equivalent to
introducing to the Einstein equations a matter source that can consist of
several different matter types -- a cosmological term, quintessence, gas of
cosmic strings, and non-relativistic cold matter.

\end{abstract}
\maketitle


The currently observed acceleration of our universe \cite{0} is the main
motivation of attempts to try to modify the theory of gravity, for example by
giving a tiny mass to the graviton. This can effectively give rise to a small
cosmological term, at least within the simplest bimetric models of massive
gravity \cite{DK}. The theory of massive gravity is not unique and there exist
a number of its models (see \cite{R} for a review). Recently a massive gravity
model was proposed which could be in some sense special, since it is the only one
that admits a quadratic action in the unbroken phase \cite{CM}. This model
reproduces in fact that in \cite{GRT} but uses a parametrization which
avoids dealing with the square root of a tensor, so that it is much simpler to
work with. In what follows we use the formulation of \cite{CM} to construct
new cosmological solutions, in particular those with spatially open and closed
metrics. It turns out that the graviton mass gives rise not only to a
cosmological term, but can also manifest itself as other matter types, with
different equations of state.

The new parametrization of the massive gravity uses the tensor \cite{CM}
\begin{equation}
S_{AB}=e_{A}^{\mu}\partial_{\mu}\phi_{B}-\eta_{AB}\,,
\end{equation}
where $\phi_{B}$ are four scalars, $\eta_{AB}$ is the Minkowski metric, the
tetrad $e_{A}^{\mu}$ determines the metric $g^{\mu\nu}=\eta^{AB}e_{A}^{\mu
}e_{B}^{\nu}$ and is constrained to fulfill the conditions
\begin{equation}
e_{A}^{\mu}\partial_{\mu}\phi_{B}=e_{B}^{\mu}\partial_{\mu}\phi_{A}\,,
\label{sym}%
\end{equation}
which insure that $S_{AB}$ is symmetric. The action can be chosen to be
\begin{equation}
S=\frac{1}{8\pi G}\int\left(  -\frac{1}{2}\,R+m^{2}\mathcal{L}\right)
\sqrt{-g}\,d^{4}x+S_{\mathrm{m}}\,,
\end{equation}
here $S_{m}$ describes the ordinary matter (for example perfect fluid), $m$ is the
graviton mass,
\begin{equation}
\mathcal{L}=\frac{1}{2}(S^{2}-S_{B}^{A}S_{A}^{B})+\frac{c_{3}}{3!}%
\,\epsilon_{MNPQ}\epsilon^{ABCQ}S_{A}^{M}S_{B}^{N}S_{C}^{P}+\frac{c_{4}}%
{4!}\,\epsilon_{MNPQ}\epsilon^{ABCD}S_{A}^{M}S_{B}^{N}S_{C}^{P}S_{D}^{Q}\,,
\end{equation}
where $S=S_{A}^{A}$ and $c_{3},c_{4}$ are two parameters (for
$c_3=c_4=0$ the action becomes quadratic). 
This theory is
equivalent to the one proposed in \cite{GRT}, where it was parameterized by
$K_{\nu}^{\mu}=\delta_{\nu}^{\mu}+\sqrt{\partial^{\mu}\phi_{A}\partial_{\nu
}\phi^{A}}$. Since $S_{AB}=-e_{A}^{\mu}e_{B}^{\nu}K_{\mu\nu}$, the two
parameterizations are equivalent, but the $S_{AB}$ parameterization has the
obvious advantage, since it does not require to take the square root which is
only defined through a series expansion.

The field equations of the above action are
\begin{equation}
G_{\mu\nu}=m^{2}T_{\mu\nu}+8\pi GT_{\mu\nu}^{\mathrm{m}} \label{ein}%
\end{equation}
with
\begin{align}
T_{\mu\nu}  &  =(S+1)e_{\mu}^{A}\partial_{\nu}\phi_{A}-\partial_{\mu}\phi
_{A}\partial_{\nu}\phi^{A}+\frac{c_{3}}{2}\,\epsilon_{ABCQ}\epsilon
^{MNPQ}e_{\mu}^{A}\partial_{\nu}\phi_{M}\,S_{N}^{B}S_{P}^{C}\nonumber\\
&  +\frac{c_{4}}{6}\,\epsilon_{ABCD}\epsilon^{MNPQ}e_{\mu}^{A}\partial_{\nu
}\phi_{M}\,S_{N}^{B}S_{P}^{C}S_{Q}^{D}-g_{\mu\nu}\mathcal{L}\,,
\end{align}
which is symmetric in view of \eqref{sym} and which has to 
satisfy  the conservation condition
\begin{equation}
\nabla^{\mu}T_{\mu\nu}=0\,,~~~~~~ \label{bian}%
\end{equation}
also $\nabla^{\mu}T_{\mu\nu}^{\mathrm{m}}=0$. Let $t,r,\vartheta
,\varphi$ be spherical coordinates and $n^{a}=(\sin\vartheta\cos\varphi
,\sin\vartheta\sin\varphi,\cos\vartheta)$ a unit vector. We choose the four
scalars to be
\begin{equation}
\phi^{0}=T(t,r),~~~~\phi^{a}=U(t,r)n^{a}\,, \label{phi}%
\end{equation}
and the tetrad
\begin{equation}
e_{\mu}^{0}dx^{\mu}=q(t,r)dt+h(t,r)dr,~~~e_{\mu}^{a}dx^{\mu}=\frac
{R(t,r)}{p(t,r)}d(p(t,r)n^{a}). \label{e}%
\end{equation}
Choosing $h=R^{2}\dot{p}p^{\prime}/(qp^{2})$ the metric becomes diagonal
\begin{equation}
ds^{2}=\left(  1-\frac{R^{2}\dot{p}^{2}}{q^{2}p^{2}}\right)  \left(
q^{2}dt^{2}-\frac{R^{2}p^{\prime2}}{p^{2}}\,dr^{2}\right)  -R^{2}%
(d\vartheta^{2}+\sin^{2}\vartheta d\varphi^{2}).
\end{equation}
Inserting \eqref{phi},\eqref{e} into \eqref{sym},\eqref{ein},\eqref{bian}, the
angular variables decouple, giving a system of non-linear PDE's for
$T(t,r),U(t,r),q(t,r),p(t,r),R(t,r)$.

Let us consider first the static case. Setting
\begin{equation}
q(t,r)=q(r),~~~R(t,r)=r,~~~p(t,r)=\exp\left(  \int\frac{N(r)}{r}\,dr \right),
\end{equation}
the metric becomes
\begin{equation}
ds^{2}=q^{2} dt^{2}-N^{2} dr^{2}-r^{2}(d\vartheta^{2}+\sin^{2}\vartheta
d\varphi^{2}).
\end{equation}
If $T^{\mathrm{m}}_{\mu\nu}=0$ then the field equations are fulfilled  by
\begin{equation}
\label{d}q^{2}=\frac{1}{N^{2}}=1-\frac{2M}{r}-\frac{m^{2}(C-1)}{3}\,r^{2}%
\end{equation}
and $T(t,r)=t$, $U(t,r)=Cr$, provided that
\begin{equation}
\label{c}c_{3}=\frac{2-C}{C-1},~~~~~c_{4}=-\frac{3-3C+C^{2}}{(C-1)^{2}},
\end{equation}
where $C,M$ are integration constants. This solution was in fact found in
\cite{Niew}, it describes the Schwarzschild-dS or (Schwarzschild-AdS) metrics.

Let us now consider time-dependent solutions. Choosing
\begin{equation}
\label{q}q(t,r)=a(t),~~~R(t,r)=a(t)\,r,~~~p(t,r)=\frac{r}{1+\sqrt{1-Kr^{2}} }\,,%
\end{equation}
with $K=0,\pm1$, the metric becomes
\begin{equation}
\label{g}ds^{2}=a^{2}(t)\left(  dt^{2}-\frac{dr^{2}}{1-Kr^{2}}-r^{2}%
(d\vartheta^{2}+\sin^{2}\vartheta d\varphi^{2}) \right)  .
\end{equation}
The matter source is chosen to be $8\pi G T^{\mathrm{m}\,\mu}_{~~~\nu
}=\mathrm{diag}(\rho(t),-P(t),-P(t),-P(t))$, whose conservation condition is
\begin{equation}
\dot{\rho}+3\,\frac{\dot{a}}{a}\,(\rho+P)=0.
\end{equation}
Choosing $U(t,r)=Cra(t)$ and
\begin{align}
K  &  =0:~~~~~T(t,r)=-\frac12\,Cr^{2}\dot{a}\,\nonumber\\
K  &  =\pm 1:~~~ T(t,r)=KC\dot{a}\sqrt{1-Kr^{2}}\,\nonumber\\
\end{align}
we find that the conservation
conditions \eqref{bian} and the symmetry conditions \eqref{sym} are fulfilled,
provided that $c_{3},c_{4}$ are again constrained according \eqref{c}, while
Einstein equations \eqref{ein} reduce to
\begin{equation}
\label{a1}3\,\frac{\dot{a}^{2}+Ka^{2}}{a^{4}}=m^{2}(C-1)+\rho.
\end{equation}
This describes the late time acceleration. Indeed, if $\rho=\gamma
P$ then $\rho\sim a^{-3-3/\gamma}$ so that for large $a$ the second term on
the right in \eqref{a1} becomes negligible. The dynamic is then driven by the
first term corresponding to the cosmological constant $m^{2}(C-1)$
which can be positive or negative, depending on value of the
integration constant $C$.
It is
worth noting that the $K=0$ cosmology was previously described in the
literature  in the decoupling limit \cite{GRT}.

One can also get solutions for generic values of $c_{3},c_{4}$. With
$U(t,r)=Cr$ and
\begin{align}                             \label{KK}
K  &  =0:~~~~~p(t,r)=r,~~~~~~~~~T(t,r)=0\nonumber\\
K  &  =-1:~~~p(t,r)=re^{t},~~~~~T(t,r)=C\sqrt{1+r^{2}}\nonumber\\
K  &  =1:~~~~~p(t,r)=re^{it},~~~~~T(t,r)=-iC\sqrt{1-r^{2}}%
\end{align}
the metric again becomes \eqref{g}, while the field equations reduce to
\begin{equation}
3\,\frac{\dot{a}^{2}+Ka^{2}}{a^{4}}=m^{2}\left(  c_{4}-4c_{3}-6+\frac
{3C(3+3c_{3}-c_{4})}{a}+\frac{3C^{2}(c_{4}-2c_{3}-1)}{a^{2}}+\frac{C^{3}%
(c_{3}-c_{4})}{a^{3}}\right)  +\rho. \label{a}%
\end{equation}
Although  the matrix $\partial_\mu\phi_A$ is degenerate for these solutions,
this should not affect the dynamics of their perturbations, so that the gravitons
should still be massive. 
We see that the value of the cosmological term $m^{2}(c_{4}-4c_{3}-6)$ is now
determined by the model parameters and not by an integration constant.
Secondly, the graviton mass also gives rise to the three additional terms in the
right hand side of the equation which decay as $1/a$, $1/a^{2}$ and $1/a^{3}$. 
These can be viewed as three different matter types contributing to the
energy density. The $1/a^{3}$ behavior of the energy is the same as for the
non-relativistic cold matter $(P=0)$, the $1/a^{2}$ is the same as for a gas
of cosmic strings, while the $1/a$ is typical for quintessence.  
We also notice that the $K=1$ solution in \eqref{KK}, obtained via analytic 
continuation of the $K=-1$ case, has complex $e_A^\mu$ and $\phi_A$, although
the metric, $S_{AB}$ and the equation \eqref{a} are real. It seems that there should
be a different representation of this solution, with real $p(t,r)$ and $T(t,r)$, but we could
not find it.

Summarizing, we have found spatially open, closed, and flat cosmologies with
massive gravitons. The graviton mass gives rise to a cosmological term thus
leading to the late time acceleration, but it can also give additional
contributions to the energy that can be viewed as matter components with
different equations of state.

A.~H.~C. is supported in part by the National Science
Foundation under Grant No. Phys-0854779.


\begin{thebibliography}{9}                                                                                                %
\bibitem {0}A.G.~Riess \textit{et al.},
\newblock {\em {\it Astron.J.}} {\bf 116} (1998) 1009; S.~Perlmutter
\textit{et al.}, \newblock {\em {\it Astrophys.J.}} {\bf 517} (1999) 565.

\bibitem {DK} T.~Damour, I.I.~Kogan,
\newblock {\em {\it Phys.Rev.}} {\bf D 66} (2002) 104025.

\bibitem {R}V.A.~Rubakov, P.G.~Tinyakov,
\newblock {\em {\it Phys.Usp.}} {\bf 51} (2008) 759; K. Hinterbichler, arXiv:1105.3735.

\bibitem {CM} A.H.~Chamseddine and V.~Mukhanov, arXiv:1106.5868.

\bibitem {GRT}C.~de~Rham, G.~Gabadadze,~A.J.~Tolley,
\newblock {\em {\it Phys.Rev.Lett.}} {\bf 106} (2011) 231101;
\newblock {\em {\it Phys.Rev.}} {\bf D 83} (2011) 103516; arXiv:1107.3820.



\bibitem {Niew}Th.M.~Nieuwenhuizen, arXiv:1103.5912.


\end{thebibliography}
\end{document}